# Apple IOS Devices for Network Administrators

Dr. Timur Mirzoev
Georgia Southern University
Statesboro, GA USA

Gerard Gingo; Mike Stawchansky; Tracy White
Southern Polytechnic State University
Marietta, GA USA

*Abstract* - As tablet devices continue to gain market share at the expense of the traditional PC, they become a more integral part of the corporate landscape. Tablets are no longer being utilized only by sales executives for presentation purposes, or as addition to the traditional laptop. Users are attempting to perform significant amounts of their daily work on tablet devices, some even abandoning the ubiquitous laptop or desktop entirely. Operating exclusively from a tablet device, specifically Apple IOS tablet devices creates unique challenges in a corporate environment traditionally dominated by Microsoft Windows operating systems. Interactions with file shares, presentation media, VPN, and remote access present barriers that users and helpdesk support are unfamiliar with in a relation to an iPad or iPhone. Many solutions are being offered to these challenges some of which are analyzed by this manuscript.

## I. INTRODUCTION

As tablet technology becomes more ubiquitous, consumers are demanding the same ease of use and mobility these provide for their business use. The limits on the devices used should not be a hindrance for the user [28].

One solution is a combination of applications, services, and hosted service modifications to best provide a seamless work experience for an end user utilizing a tablet. The app store provides many options for enhancing the corporate access experience. Many corporate systems can be modified to allow for easier access by tablet and mobile devices. Some systems will require upgrades while others may require third party modules to enable this additional functionality [26].Consolidation of the best methodologies into a concise reference for administrators and users will facilitate integration of consumer tablet devices into the corporate enterprise.

Modern network administrators today can be tasked with maintaining very large networks with data centers and hubs located in multiple locations. Businesses are looking to try to decrease the number of IT staff, and creatively find ways to increase productivity for existing staff. In order to help facilitate the increased demand for productivity and availability from an administrator, the use of iOS tablets and phones from a network administration standpoint can be very useful. Tablets in the form of iPads and even iPhones are becoming important tools that are the key tools for a network administrator. VPN and remote desktop capability provides them with a way to perform their tasks from remote locations.

There are many advantages to incorporating tablets running iOS into an information technology infrastructure [2].Tablets are small and easy to handle, usually less expensive than PCs with a longer battery life, with the ability to record information directly on the tablet. Businesses small and large, as well government agencies can recognize how this technology can be beneficial in their daily operations especially for duties performed outside the office. Professions that require work outside the normal office setting appreciate the flexibility provided by a device like a tablet [21].Understanding the dynamics of the tablet user in the course of business is important when transitioning from traditional technology like desktops or laptops. The interface of the applications used must be succinct and direct in order to experience the best use of the tablet.

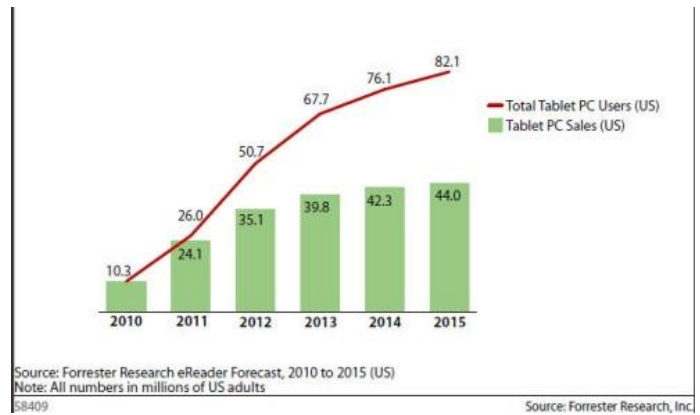

Figure 1: Tablet sales and use [7]

The research will show that iOS devices have a place in administering support to all possible users and providing the right integration will make the transition to these devices successful.

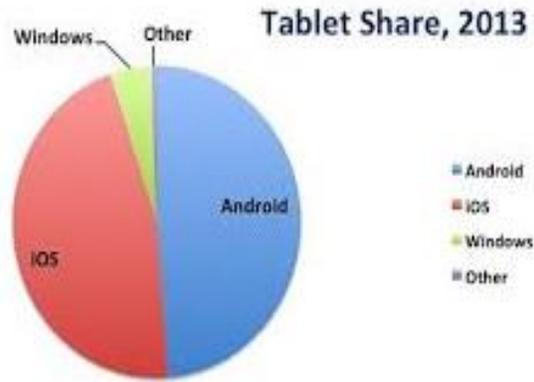

Figure 2: Tablet shares among various operating systems [20]

While the network administrator performs various technical tasks, the day-to-day business user will be interacting with systems in a broader scope. Corporate file shares can pose a challenge for users, particularly those tailored for Windows Operating systems. As access devices become more diverse, helpdesks struggle to provide a unified access methodology [24]. Identifying solutions that allow users to access, modify, and share documents stored on corporate file shares solely using their tablet devices will give users the freedom to choose their access device rather than requiring the traditional laptop.

As the convergence of consumer devices and corporate networks accelerates, as IT professionals we must be ready to meet requests to utilize these resources with non-standard devices [18].Users no longer wish to segregate their technology into consumer devices and corporate devices. Their level of technical sophistication is increasing rapidly, as is their demand for corporate technology to function as seamlessly as their consumer-focused technology. While this creates unique challenges, it also offer opportunities to provide an enhanced user experience through the consumer focused Apple iOS device (Clevenger, 2011).

## II. EMERGING TECHNOLOGY

Today, there are many jobs requiring employees to perform work-related duties outside the normal office setting. Businesses will need to consider employees' need to access files and applications while mobile.  In a recent survey conducted by [5], 610 respondents said tablet use is critical to their daily work routine and even provides for better productivity [5].  The respondents were from varying businesses and professions.  Not only there is a need for the use of tablet technology using various operating systems but also there is a need to make the interface easier for all technology.

Another interesting trend is the Bring Your Own Device (BYOD) program [17]. In an effort to alleviate the need for businesses to purchase tablets for employees, some businesses are considering employees use their own devices to access the company network.  Allowing the employee to use their personal tablet is not only productive but saves the company money [15], [7]. Like with any technology, there are multiple concerns facing businesses that would like to accommodate their employees on the go.  Among these concerns, the normal security issues but more importantly the issue of compatibility [9].With the market increase of tablets like iPads and smartphones like iPhones, compatibility clearly impedes the ability for the two technologies to work together.  The iOS technology is not available for many businesses and this is one of the obstacles facing users.  The operating system is the main reason many employees that use iPads or iPhones are limited in their ability to obtain access to a business network.

There are solutions to the challenges facing companies and its employees that will allow them to move to easier integration of the technology [12].The important aspect of this research is to demonstrate that the solutions are viable and expanding technology will allow network administrators the ability to utilize iOS technology [17].

One of the largest hurdles for enterprises considering iOS device integration is the ability of network administrators to utilize and manage the devices. However, as the demand is quickly growing with companies adopting iOS devices, companies are finding new ways to meet demand. A case study done interviewing Genetech, a bio-tech giant, implemented an enterprise app store custom built for themselves. This custom store eased and streamlined adoption of iOS devices. Private app stores allow companies to help employees working in BYOD programs find the apps they need quickly and easily. By providing a list of approved or custom built apps, it makes it easier not only from an administration standpoint, but also from a user feedback perspective. Using these methods allowed for a deployment that provided effective, sustainable results [19].

For applications that may not be readily portable to the iOS, there is still the option of remote desktop capabilities. For example, network administrators may have specialized tools that they can access on their regular laptops, desktops, or servers themselves that are not available on iOS devices. A network administrator may need to access proprietary GUI interfaces built to be accessed server side. . An example of such a product is VMWare's consoles or Microsoft's Active Directory control panel.. In these cases, several choices exist for the iPad that make use of the Windows remote desktop protocol, or the popular VNC software [4].

As network administrators should be concerned about security, iOS devices are able to support varying methods of security. By far, one of the most important technology is the ability to connect to a VPN, which is required for many types of connections to either transfer data or remotely administer a network. People still argue whether if iOS devices are really enterprise security ready [16]. With the popularity of iOS devices increasing, more built-in security, compatibility, and supporting apps are being natively supported in the iOS. Cisco, a major player in network security, now has worked with Apple to include a stable iOS VPN client that meets enterprise security standards.

Multiple options are available to ease access to shared storage for Apple iOS devices. Some products are in the early stages of availability for enterprises, such as VMware's Horizon Suite. The Horizon Suite is a rebranded Project Octopus that VMware has promoted for several years as a method to unify data and application access in the post-PC era [30]. Other products have been under development by startup companies that recognized the need for easy access to data across platforms. Syncplicity is a start-up that was acquired by EMC in 2012. Syncplicity's initial feature set allowing for synchronization of data from the local network to cloud networks has been quickly bolstered by EMC's internal development team to expand feature sets and integration with their VNX storage platforms. This offering provides a similar user experience whether on laptop, IPhone, or iPad device [22]. Such an investment by the leading storage provider highlights the perceived need in the marketplace for consumer focused products. Oxygen Cloud is a product that can partner with EMC and other storage solutions to provide Dropbox style functionality. Dropbox is cloud based storage service that stores and shares documents between multiple devices and users. Oxygen Cloud mimics Dropbox functionality, but utilizes corporate resources rather than a third party service. Oxygen Cloud touts an enterprise class product with consumer friendly interfaces for all devices. Integration of this product into an existing SMB file sharing platform appears to provide ease of access to multiple mobile device styles including Apple iOS devices [26]. Oxygen Cloud, VMware Horizon, and Syncplicity all provide the end user with an enhanced experience in regards to utilizing corporate file shares. These file sharing technologies combined with the other applications and services can provide a foundation for an integrated mobile environment. While these solutions require an investment in capital and resource time to configure, they are end-user focused from the client perspective. This focus will ultimately drive acceptance of the technology by Information Technology departments that are struggling to deal with consumer device limitations.

III. PRACTICAL APPLICATION

Oxygen Cloud, VMware Horizon, and Syncplicity all have compelling functionality that provides differing options to integrate Apple IOS devices into networks. It appears that Oxygen Cloud offers the broadest support in regards to infrastructure that can be utilized [26]. The wide range of infrastructure that can be utilized is from Windows and Linux file shares, to cloud based storage providers such as Amazon S3 or IBM SmartCloud. In order to minimize the costs associated with a lab environment with which to test Oxygen Cloud, a free file share technology, called FreeNAS, will be utilized in a virtual environment.

FreeNAS is an open source project that provides enterprise level file sharing infrastructure with minimal overhead on a multitude of platforms. FreeNAS supports SMB/CIFS, NFS, and AFP file access methods allowing for the simulation of multiple infrastructure sources [14]. Each source type that is supported by Oxygen Cloud is utilized to present file shares to various Apple IOS devices.

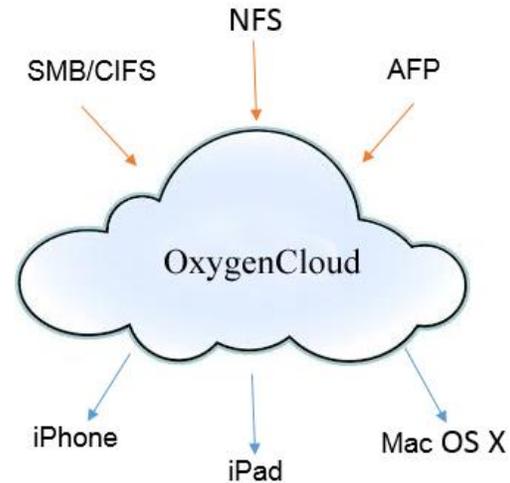

Figure 3: File systems and device types

The devices tested include an IPhone 4, IPhone 4S, IPAD 2, and MacBook OS X laptops. In the testing for this study, a combination of source file systems with varying Apple IOS devices the test scenarios may present the effectiveness of the product in providing both simplified access and administration.

The test scenarios include the following:
- Ability to attach to a shared storage
- Ability to view, create, and modify documents on Oxygen Cloud shared storage
- Ease of access to Oxygen Cloud shared storage on a scale of 1 – 5 compared to standard share
- Ease of configuration of Oxygen Cloud shared storage

These test scenarios allow for a wide variety of iOS devices to be benchmarked. A large test footprint is needed in order to support a BYOD environment where multiple iOS devices could be utilized on a daily basis.

Aside from a user wishing to pull data from network storage, there is a need to control software not readily available in a public or corporate app store. "Jump" is an iOS app that is currently well rated in Apple's App store and supports all existing RDP and VNC functionality using just an IP address or host name. For those concerned with security, Jump works with secure RDP, TLS/SSL, and FIPS layers that meet standard security requirements for corporate networks [11]. Utilizing Jump software to perform remote control operations can streamline administration without a larger laptop. A test for the usability of this application can be conducted with an end-user survey of system administrators. Marks can be given for interface, compatibility with different server operating systems, and ease of use on different iOS devices in scales of 1-5.
- Tests for each iOS device type to be evaluated:
    - Ability to connect by IP and Host Name to each server operating system type

- o Testing interface behavior between remote operating systems
- o Perform variety of daily tasks using Jump
- Measuring satisfaction for each iOS device type for overall user experience:
    - o Rate the compatibility of the app across devices
    - o Rate the usability of the app overall
    - o Rate the overall experience using the app

This test measures the effectiveness of how well Jump works in a variety of situations and can provide valuable analytics.

It is evident that there is a need to consider the current trend in tablet technology used by network administrators. These tablets run not only the usual and customary operating systems, but iOS devices as well. Furthermore, having the option of using a tablet running iOS provides a way to support the business network when mobility is required. Providing the right integration makes the transition to mobile devices successful.

According to Hess in the article "10 BYOD Mobile Device Management Suites You Need to Know", "the single most important qualifying or disqualifying point is that the mobile device management or MDM software must support more than a single platform. The second is that it must support Apple's iOS". This is important since iOS devices have a large market share [25].

In another study conducted by Frost and Sullivan (2012), businesses are beginning to implement some type of mobile app to interface with the different devices used by its employees. This further supports the importance of increasing the technology needed to support devices running technology like iOS. According to the study, approximately 82% of the businesses surveyed have some type of application established.

A survey conducted consisting of a small sampling of 12 network administrators working in the Information Systems department at DeKalb County Government showed 8 out of 12 have some type of personal device running iOS technology [31].

The three survey questions presented to the respondents were:

1. Do you personally own a device that is running iOS technology?
2. Are you able to use this device to provide network support?
3. Would you like to use your personal device that is running the iOS technology?

Additional results obtained from this survey showed iOS devices could not access the network because the network is only compatible to devices that work with Windows. With further discussion, it was learned that due to budgetary constraints, there were only four devices capable of accessing the network and were being shared by the 12 administrators. This survey provided further evidence that if technology was implemented to allow devices running iOS technology, each individual administrator would have the ability to utilize personal devices. This could be advantageous for the business.

Various methods like research, surveys and testing will support the case for compatible technology to be all inclusive of different devices. Removing the restrictions to which devices are used will help network administrators be better equipped to provide the needed support with the ability to utilize personal devices. The findings presented through actual testing will substantiation the research and assist with confirmation that the technology is capable of allowing network administrators using iOS devices a true position in the area of network support.

IV. FINDINGS

Oxygen Cloud testing requires several pieces of software to start the process. A FreeNAS instance following the community guidelines for establishing a virtual FreeNAS environment is utilized as the source file share [13]. A baseline virtual machine consisting of 512MB or RAM and two 10GB virtual disks to support the FreeNAS data shares is configured. Once created, the virtual disks are utilized to create a CIFS share that was presented to the network. Oxygen Cloud requires an Oxygen Storage Connector in order to present this CIFS share to various clients. A demo license of this is available from Oxygen Cloud directly. The Oxygen Cloud Storage connector is a virtual appliance that is running in the same network segment that the FreeNAS CIFS share resides. While Oxygen Cloud utilizes internal storage, it has a portion that is deemed SaaS or Software-as-a-Service. The Oxygen Cloud virtual appliance connects to this SaaS that facilitates the utilization of the backend storage [26]. While the actual configuration of the virtual appliance is not onerous, there is a certain amount of prerequisite work requiring firewall modifications, SSL certificates, and DNS additions. After this configuration portion is complete and a temporary SaaS account established, the client portions are addressed.

As the test devices are an IPad 2, IPhone 4s, and a MacBook Pro OS X laptop this requires three differing Oxygen Cloud client installs. The mobile devices utilize Oxygen Mobile for iPad and Oxygen Mobile for iPhone from the iTunes store respectively. The MacBook Pro utilizes a DMG file for installation that is downloaded directly from the Oxygen Cloud website. Once each device completes installation, it simply requires an Oxygen ID to connect to the Oxygen Cloud and be presented with storage to be utilized. For the mobile devices, this access is through the Oxygen application. Access is relatively easy, and files modified on the device are able to be shared amongst all devices. The MacBook Pro has the additional feature of having a designated Oxygen drive. The Oxygen drive functions as a normal file folder with the exception that files placed in it are available on the mobile devices in approximately 10 seconds. There is also functionality integrated to share files between multiple user groups, or send links containing these files through email. Testing of this functionality showed that it functioned as expected with the devices requiring minimal configuration by the end user in order to utilize Oxygen Cloud.

The test for Jump included using an iPad 2 and an iPhone 5, and using them to remotely connect to both Windows Server 2003 and Windows Server 2008 r2 over a corporate VPN. A good feature of the Jump application is that it does not need any additional setup to be performed on the target server, as it uses the standard RDP protocol, so setup was not an issue. Once connected into a network through VPN, both devices had no problem connecting to either operating system via IP address. Once connected to the servers, the interface is very familiar to use on both devices. Though the screens on the iOS devices are smaller, navigation is possible by using pinch-to-zoom and scrolling the screen around to view the entire server desktop. Use is cramped on the iPhone, but is plenty adequate for simple tasks that include unlocking accounts or restarting services. With the iPad's larger screen, it was easier to run applications and interact with them on the remote system. Using more user input intensive, business applications including Microsoft Excel are possible using the iPad's on screen keyboard. From a standpoint of overall experience, the Jump application is quite successful in terms of allowing someone to work free of their laptop or workstation, and can allow a network administrator to accomplish their daily tasks.

It is apparent that there is a need to provide a compatible interface allowing administrators the ability to connect to the business network through such devices. The key findings are supported by the newly created applications to help implement solutions for network administrators using a device running iOS technology. It is clearly an issue being explored and resolved as more software developers are creating applications for running iOS technology. Through the research, the impact of iOS technology is definitely emerging and expanding and more importantly, affirms that iOS technology is gaining more attention.

In summary, the overall results include:

- Network Administrators are eager to use devices operating with non-standard technology. If given the option to utilize different technology, users will welcome the opportunity.
- Companies are supporting users of iOS devices by providing non-standard technology. Various applications are available and are being implemented to work with iOS devices. No longer do limitations on compatibility discourage users.
- Network Administrators who have the opportunity to use iOS supported technology are comfortable with participating in bring your own device or BYOD programs. Companies are open to the idea of a BYOD and are prepared to incorporate the right technology to allow the program to be successful [32]. Too many limitations like security concerns will not prevent the incorporation of this program. Most applications are addressing and resolving security issues while the devices come complete with security measures in place.
- By giving network administrators access to a broader range of resources and technologies, productivity can increase [3].
- There are many different solutions available now and more are becoming available. Businesses are beginning to take advantage of this opportunity.
- Availability of software components such as Oxygen Cloud and FreeNAS can provide a seamless integration.

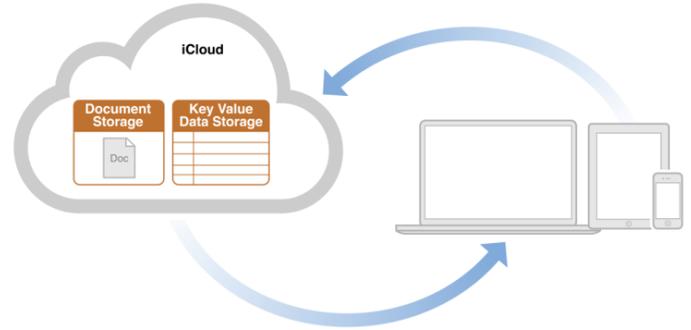

Figure 4: The figure shows the integration of iCloud document storage.

V. CONCLUSIONS

In conclusion, it has been demonstrated that significant progress has been made towards an equilibrium where mobile device convenience and administrator supportability meet. There are multiple options to achieve this end state. There several presented here are one pathway to expanding the role of iOS technology in the corporate environment in a responsible manner. Implementation of these examined technologies, or a similar suite of utilities, can provide a secure environment for users to access corporate resources with familiar and accessible consumer devices.

End users are seeking the same ease of use and mobility tablets provide for their business use. A solution to the problem is a combination of applications, services, and hosted service modifications to best provide a seamless work experience for an end user utilizing a tablet.

The practical applications performed for this research of various devices running iOS technology lends support to the idea that all technology is feasible. With the ability to allow the integration of mobile devices like iPads or iPhones into the corporate structure, it amplifies the end user's desire for ease of use and mobility enhanced. The goal of this research was not to support a complete reliance on iOS technology or to persuade one to believe any one technology is better. The goal was to show that there are options for network administrators to utilize. As with many decisions about things used in the world today, like choices in automobiles or grocery stores, it is entirely up to the user whether the technology works for the given situation. The strengths and weaknesses of any chosen technology and the ability to have a choice are based on the individual user and the situation presented.

AUTHORS PROFILE

**Dr. Timur Mirzoev** is an Associate Professor of Information Technology Department at Georgia Southern University, Allen E. Paulson College of Engineering and Information Technology. Dr. Mirzoev heads the Cloud Computing Research Laboratory, Regional VMware IT Academy and EMC Academic Alliance at Georgia Southern University. Some of Dr. Mirzoev's research interests include server and network storage virtualization, cloud systems, storage networks and topologies. Currently, Dr. Mirzoev is holds the following certifications: VMware Certified Instructor, VMware Certified Professional 5, EMC Proven Professional, LefthandNetworks (HP) SAN/iQ, A+.

**Gerard Gingco** is currently pursuing a BS in Information Technology at Southern Polytechnic State University, he works in the engineering and consulting fields with experience in the federal, banking, and healthcare information technology industries.

**Mike Stawchansky** is currently pursuing a BS in Information Technology at Southern Polytechnic State University. He is currently the Sernior Director for Systems Operations at WebMD.com. Interests include virtualization ,public/private cloud hybrids, and all manner of IT infrastructure.

**Tracy White** was born in Waterbury, Connecticut 1961, is seeking a BS degree in Information Technology at Southern Polytechnic State University. She is currently an Administrative Manager for DeKalb County Government Property Appraisal. Her interests include web design and end-user desktop support.